\documentstyle[twocolumn,aps,epsfig,floats]{revtex}

\draft 

\begin{document} 
\twocolumn[\hsize\textwidth\columnwidth\hsize\csname @twocolumnfalse\endcsname

\title{ Charge and Orbital Ordering and Spin State Transition \\
        Driven by Structural Distortion in YBaCo$_2$O$_5$ } 
\author{ S. K. Kwon and B. I. Min } 
\address{ Department of Physics, 
          Pohang University of Science and Technology, 
          Pohang 790-784, Korea } 
\date{\today}
 
\maketitle 
 
\begin{abstract}    
We have investigated electronic structures 
of antiferromagnetic YBaCo$_2$O$_5$ 
using the local spin-density approximation (LSDA) + $U$ method.
The charge and orbital ordered insulating ground state is 
correctly obtained with the strong on-site Coulomb interaction.
Co$^{2+}$ and Co$^{3+}$ ions are found to be in the high spin (HS)
and intermediate spin (IS) state, respectively.
It is considered that the tetragonal to orthorhombic structural transition 
is responsible for the ordering phenomena and the spin states 
of Co ions. The large contribution of the orbital moment 
to the total magnetic moment indicates 
that the spin-orbit coupling is also important in YBaCo$_2$O$_5$.
\end{abstract}

]

\narrowtext  

Recently, an interesting spin state transition of the Co$^{2+}$ ion 
in YBaCo$_2$O$_5$ has been reported by Vogt {\it et al.}\cite{Vogt}
using the neutron powder diffraction (NPD) measurements. 
The transition is induced by the long-range orbital 
and charge ordering of Co$^{2+}$/Co$^{3+}$ ions.
The ordered oxygen-deficient double perovskite 
$R$BaCo$_2$O$_{5+\delta}$ ($R$ = rare-earths)\cite{Zhou} 
has attracted much attention as a new spin-charge-orbital 
coupled system like manganites and also as a new Co-based 
colossal magnetoresistance (CMR) material. 
Indeed, giant magnetoresistance are observed for $R$=Gd and Eu,
$(\rho_0-\rho_{H = 7\rm T})/\rho_0$ = 41\% and 40\% 
for GdBaCo$_2$O$_{5.4}$ and EuBaCo$_2$O$_{5.4}$, respectively\cite{Martin}.

In the paramagnetic phase, YBaCo$_2$O$_5$ is crystallized 
in the tetragonal structure of the space group $P4/mmm$.
It consists of double CoO$_5$ square base pyramidal 
backbone layers along the $c$-axis 
in which Y and Ba layers intervene alternatively 
and oxygens are deficient exclusively 
from the Y layers\cite{Zhou,Martin,Maignan}.
According to the valency consideration, Co$^{2+}$ and Co$^{3+}$ ions coexist
similarly as the Mn$^{3+}$/Mn$^{4+}$ covalency 
in hole-doped La$_{1-x}$Sr$_{x}$MnO$_3$.
Below $T_{\rm N} \sim 330$ K, YBaCo$_2$O$_5$ undergoes 
a $G$-type antiferromagnetic (AFM) transition 
and the lattice changes slightly 
from the tetragonal to orthorhombic structure\cite{Vogt}. 
At $T_{\rm CO} \sim 220$ K, a pronounced upturn is observed 
in the resistivity indicating that another transition takes place, 
{\it i.e.}, the long-range charge and orbital ordering of
Co$^{2+}$/Co$^{3+}$ ions.
The stripe type charge ordering is formed in the $ab$ plane.
Further, the spin state of the Co$^{2+}$ ion changes from the
low to high spin
across $T_{\rm CO}$, which is evidenced by the increased magnetic moment 
per Co ion from $2.10 \mu_B$ at 300 K to $3.45 \mu_B$ at 25 K.
More recently, for the isostructural HoBaCo$_2$O$_5$,
essentially the same features are observed of $T_{\rm N} \sim 340$ K 
and $T_{\rm CO} \sim 210$ K\cite{Suard}.

The phenomenon of the spin state transition is observed usually 
in cobaltates such as LaCoO$_3$ and La$_{1-x}$Sr$_x$CoO$_3$. 
In LaCoO$_3$, the magnetic ground state of Co$^{3+}$ ion corresponds 
to the low spin (LS) state with $t_{2g}^6e_{g}^0$.
Upon heating, a successive spin state transition to an intermediate
spin (IS) state and then to a high spin (HS) state occurs\cite{Korotin}.
Note that, distinctly from the case in LaCoO$_3$, the spin state transition 
in YBaCo$_2$O$_5$ seems to occur for Co$^{2+}$ ion. 
In fact, this issue is still controversial.
Based on the reduction of the magnetic susceptibility below 220K,
Akahoshi and Ueda\cite{Akahoshi} suggested that the AFM transition 
takes place at $T \sim 220$ K which is associated with a spin state 
transition of Co$^{3+}$ from the HS to LS state upon cooling.
Thus, the nature of the spin state transition and the interplay
of the spin state with the charge and orbital ordering 
are still unclear.

To reveal the mechanism of the spin state transition 
as well as the charge and orbital ordering,
we have explored the electronic structure of the $G$-type AFM YBaCo$_2$O$_5$ 
using the local-spin density approximation (LSDA) + $U$ scheme
implemented in the linearized muffin-tin orbital band 
method\cite{Kwon1,Liechtenstein}.
 
We have employed two structural data of 
nearly tetragonal structure at 300 K (L1) 
and orthorhombic structure at 25 K (L2)\cite{Vogt}.  
In the L1 structure, all the Co sites have 
an equal average bond length of $d$(Co-O) = 1.97 \AA.
Whereas in the L2 structure, there are two different kinds of Co sites, 
CoI and CoII. The bond length at CoI sites is extended to 
$d$(CoI-O) = 2.03 \AA~ and that at CoII sites becomes shortened
to $d$(CoII-O) = 1.92 \AA. 
The $G$-type AFM spin order is assumed in all our calculations.

\begin{figure}[t]
\epsfig{file=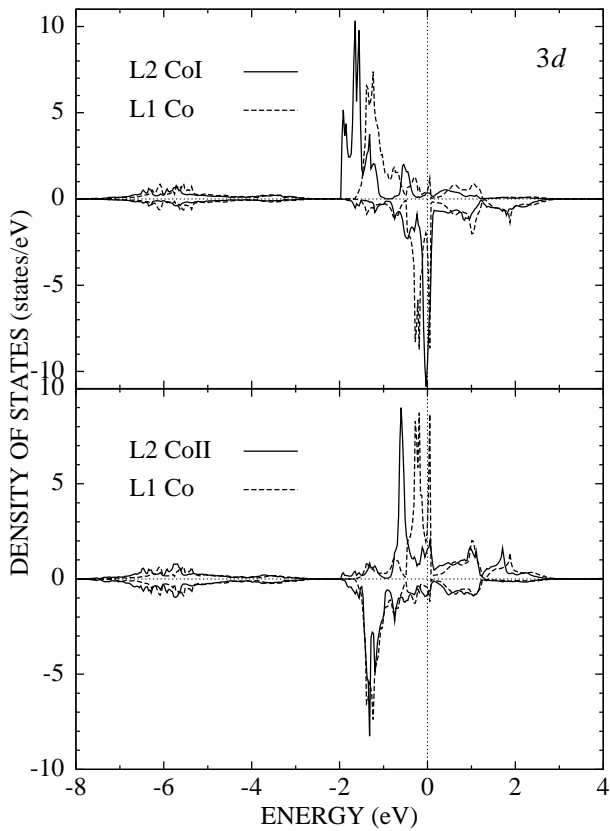,width=8.6cm}
\caption{\label{fig1} 
The LSDA PDOS of Co $3d$ electrons in the L1 (300 K phase) and L2 (25 K phase) 
structures. In L2, there are two different types of Co sites, CoI and CoII.
Neighboring CoI and CoII sites are polarized antiferromagnetically.
As manifested by the exchange splitting, 
$3d$ electrons are more localized for CoI site and less localized for CoII
than those for Co in L1.
}
\end{figure}

In Fig. \ref{fig1}, we have compared the LSDA 
Co $3d$ partial density of states (PDOS) for the L1 and L2 structures.
Most notable is the change of the exchange splitting 
which becomes larger for CoI site and smaller for CoII in L2
than that for Co in L1. 
This indicates that, due to the lattice distortion, 
$3d$ electrons in L2 becomes more localized for CoI 
and less localized for CoII in comparison to those for Co in L1.
The calculated spin magnetic moments are $\mu_S = 1.68 \mu_B$ 
and $0.95 \mu_B$ for each CoI and CoII site, respectively, in L2
and $\mu_S = 1.13 \mu_B$ for Co in L1.
The sizes of the spin moments are consistent 
with the degree of $3d$ electron localization.
Hence, the LSDA gives a qualitatively good information
about the structural transition effects on Co $3d$ electron states.
The LSDA, however, yields an incorrect metallic ground state
which pertains even after the structural transition 
from the L1 to L2 structure.
This is different from the experiment which shows unambiguously
the semiconducting resistivity behavior below 220 K\cite{Vogt,Suard,Akahoshi}. 

Using the LSDA, one cannot expect proper description of localized Co $3d$ 
electrons in YBaCo$_2$O$_5$.
To resolve the above problem, we have applied the LSDA + $U$ method 
with parameter values of $U = 5.0$ eV and $J = 0.89$ eV. 
Although there is an arbitrariness of the $U$-value in our calculation,
the LSDA + $U$ results are usually not much sensitive to the used $U$-value 
within $\Delta U \sim \pm 1$ eV\cite{Kwon2}.
The spin-orbit interaction is also taken into account
in the self-consistent variational loop,
because Co $3d$ electrons are expected to retain atomic properties 
to some extent due to their localized nature. 

\begin{figure}[t]
\epsfig{file=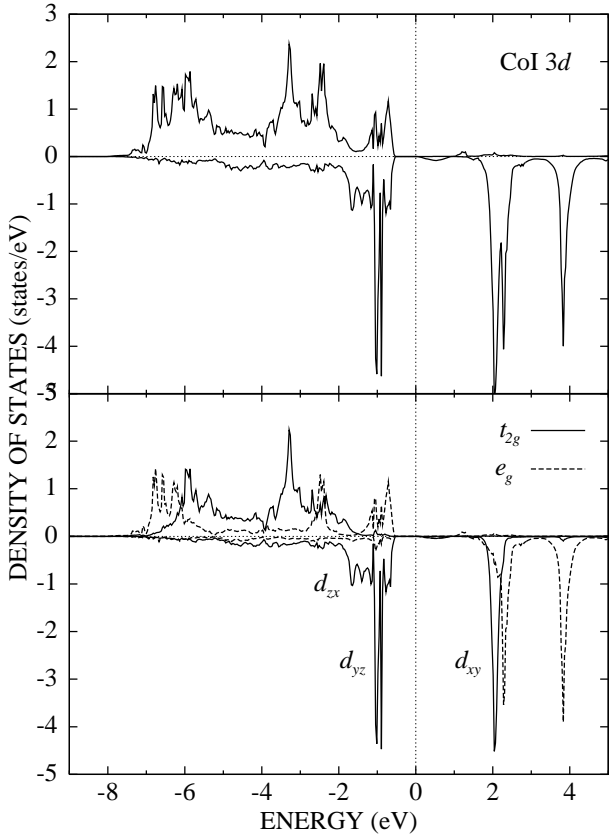,width=8.6cm}
\caption{\label{fig2}
The LSDA + $U$ PDOS of CoI $3d$ electrons in the L2 (25 K phase) structure. 
In the bottom panel, $t_{2g}$ and $e_g$ decompositions of
$3d$ PDOS are also provided.
The majority-spin bands are fully occupied,
while in the minority-spin bands, only 2/3 states of $t_{2g}$ bands 
are occupied.
The nominal valency of CoI site is Co$^{2+}$ with $3d^7$ ($t_{2g}^5e_{g}^2$).
Co$^{2+}$ ion is in the HS state.
}
\end{figure}

\begin{figure}[t]
\epsfig{file=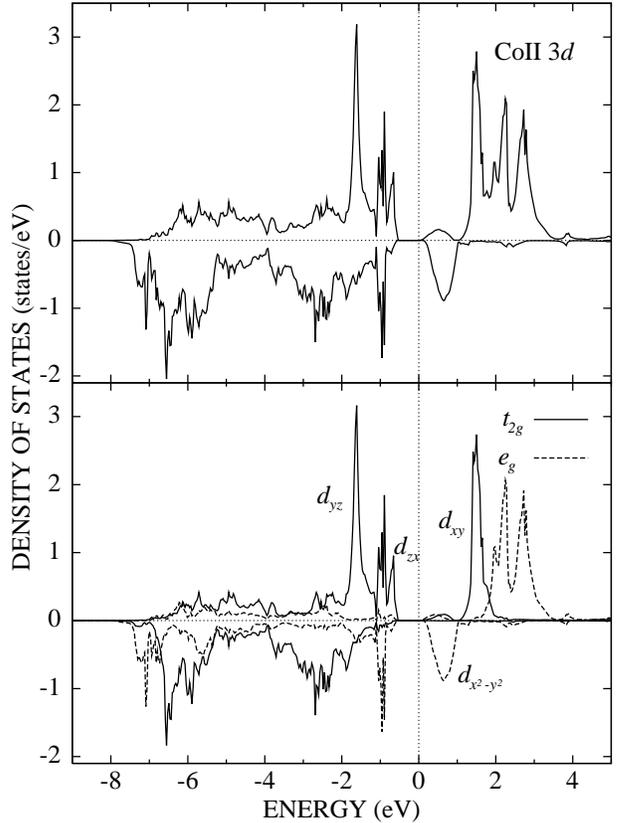,width=8.6cm}
\caption{\label{fig3}
The LSDA + $U$ PDOS of CoII $3d$ electrons in the L2 (25 K phase) structure
which is antiferromagnetically polarized to that of neighboring CoI. 
Unoccupied $e_g$ states in the majority-spin bands are clearly visible 
above $E_{\rm F}$.
The valency and the $3d$ electron configuration of CoII site is 
nominally identified as Co$^{3+}$ and $3d^6$ ($t_{2g}^5e_{g}^1$), 
respectively.  Co$^{3+}$ ion is in the IS state.
}
\end{figure}

Figures \ref{fig2} and \ref{fig3} present CoI and CoII $3d$ PDOS, 
respectively, in the L2 structure obtained by the LSDA + $U$ calculations. 
In the bottom panels, $t_{2g}$ and $e_g$  decompositions of
$3d$ PDOS are also provided.
It is amusing to note that the energy gap of $E_g \sim 0.6$ eV opens 
at the Fermi level $E_{\rm F}$ and so YBaCo$_2$O$_5$ becomes an insulator
as expected in consideration of the Co$^{2+}$/Co$^{3+}$ charge ordering.
For CoI $3d$ electrons, the majority-spin bands are fully occupied 
by three electrons in $t_{2g}$ and two electrons in $e_g$ bands. 
The minority-spin $t_{2g}$ bands are only partially occupied 
by two electrons, and one $t_{2g}$ and two $e_g$ bands are almost empty. 
Considering the pyramidal environment of Co, 
the occupied $t_{2g}$ states corresponds to $d_{zx}$ and $d_{yz}$ 
while the empty $t_{2g}$ to $d_{xy}$.
Accordingly, the nominal valency and the $3d$ electron configuration 
at CoI site are assigned to be Co$^{2+}$ and $3d^7$ ($t_{2g}^5e_{g}^2$), 
respectively\cite{Nominal}.  
Hence, Co$^{2+}$ ion is in the HS state with spin magnetic moment 
of $\mu_S = 3 \mu_B$ ($S = 3/2$), which is consistent with the NPD data.

For CoII $3d$ electrons, the majority-spin bands are not fully
occupied with a split-off empty $e_g$ state 
above $E_{\rm F}$ (see Fig. \ref{fig3}). 
For the minority-spin bands, the situation is similar to the case of CoI.
It is thus possible to identify the $3d$ electron configuration of CoII
as $3d^6$ ($t_{2g}^5e_{g}^1$) and the valency as Co$^{3+}$.
With one less electron than Co$^{2+}$, the lower $d_{3z^2-r^2}$ 
out of two $e_g$ states in the majority-spin bands is occupied
and the upper $d_{x^2-y^2}$ becomes empty.
Hence, the spin state of Co$^{3+}$ ion is the IS state
which is also in agreement with the experimental analysis 
of $\mu_S = 2 \mu_B$ ($S = 1$).

The calculated charge occupancies of each Co $3d$ orbitals 
are shown in Table \ref{table1}. 
For Co$^{2+}$ ion, the majority-spin bands are almost completely occupied 
by $t_{2g} \uparrow = 2.98$ and $e_{g} \uparrow = 1.99$ electrons,  
while the minority-bands are only partially occupied by 
$t_{2g} \downarrow = 2.02$ 
and $e_{g} \downarrow = 0.37$ electrons.  
For Co$^{3+}$ ion, it is noticeable that the majority-spin $e_{g}$ states 
are partially occupied by $e_{g} \downarrow = 1.54$ electrons.
Hence, the calculated charge occupancies are consistent with 
the nominal valencies of Co$^{2+}$ and Co$^{3+}$, 
if one takes into account the band hybridization effects.

\begin{table}[b]
\caption{\label{table1} The calculated charge occupancies 
of Co $3d$ orbitals (electrons) and magnetic moments ($\mu_B$) 
of YBaCo$_2$O$_5$ in the LSDA + $U$ method.}
\begin{tabular}{lcl} 
                   & ~Co$^{2+}$          &  Co$^{3+}$    \\ \hline
$t_{2g} \uparrow$  &  2.98               &  2.14         \\  
$e_{g} ~\uparrow$  &  1.99               &  0.60         \\  
$t_{2g}\downarrow$ &  2.02               &  2.99         \\ 
$e_{g}~\downarrow$ &  0.37               &  1.54         \\ \hline  
$\mu_S$            &  2.61               &  1.84         \\
$\mu_L$            &  1.04               &  0.40         \\ 
$\mu_{\rm tot}$    &  3.65               &  2.24         \\ 
$\mu_{\rm exp}$    &  4.2$^a$,3.7$^b$    &  2.7$^{a,b}$  \\
\end{tabular}
$^a$ Reference\cite{Vogt}. \\
$^b$ HoBaCo$_2$O$_5$ in Ref.\cite{Suard}.  
     The assignment of the magnetic moments to each Co ion is corrected
     (see text). \\
\end{table}

Vogt {\it et al.}\cite{Vogt} have deduced magnetic moments 
from the NPD experiments as $\mu_{\rm exp} = 4.2 \mu_B$ and $2.7 \mu_B$
for each Co$^{2+}$ and Co$^{3+}$ ion, respectively.
In the analysis of the experiments, they counted only the spin moment
contribution, assuming that the orbital moment is quenched.
However, the orbital moment is only partially quenched in YBaCo$_2$O$_5$.
In Table \ref{table1}, we have summarized 
the calculated magnetic moments using the LSDA + $U$ method.
For Co$^{2+}$ ion, the spin and orbital magnetic moments are 
$\mu_S = 2.61 \mu_B$ and $\mu_L = 1.04 \mu_B$, respectively, 
and for Co$^{3+}$ ion, $\mu_S = 1.84 \mu_B$ and $\mu_L = 0.40 \mu_B$.
The orbital moment of Co$^{2+}$ ion is 
as much as that of CoO\cite{Svane,Solovyev}, 
and the orbital moment of Co$^{3+}$
is comparable to that of NiO\cite{Kwon2,Fernandez}.
The non-negligible orbital moment, which originates from the localized nature 
of Co $3d$ electrons, suggests that YBaCo$_2$O$_5$ should fall 
in a class of strongly correlated electron system like CoO and NiO.
The calculated total magnetic moments of $\mu_{\rm tot} = 3.65 \mu_B$ 
and $2.24 \mu_B$ for Co$^{2+}$ and Co$^{3+}$ ion, respectively, are 
only slightly smaller than the experimental values.
Evidently, this interpretation will also be valid for HoBaCo$_2$O$_5$.
Suard {\it et al.}\cite{Suard} have improperly assigned 
the NPD measured $\mu_{\rm exp} = 3.7 \mu_B$ and $2.7 \mu_B$
in HoBaCo$_2$O$_5$ to Co$^{3+}$ and Co$^{2+}$ ion, respectively,
However, their assumptions of spin-only moments and the HS states for 
both Co$^{2+}$ and Co$^{3+}$ ions are discarded by the present results.
To test the calculated results, more experimental works 
like the x-ray scattering measurement are encouraged, 
in which separate determination of the spin and orbital moments are possible. 

\begin{figure}[t]
\epsfig{file=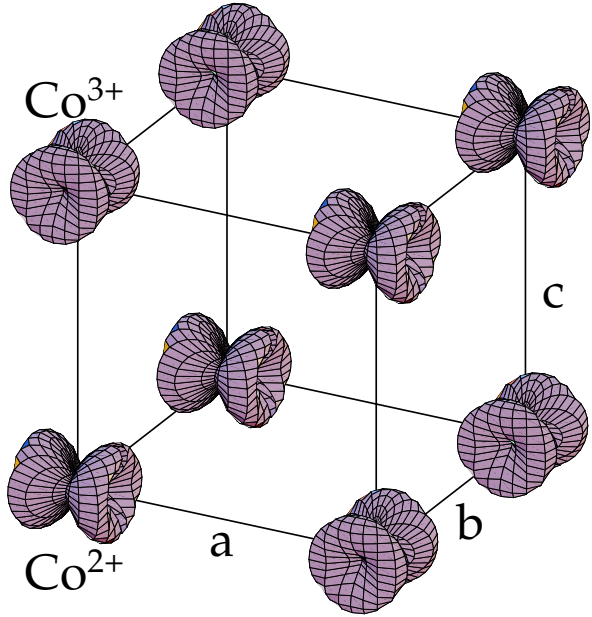,width=8.6cm}
\caption{\label{fig4} The orbital ordering geometry of
the occupied minority-spin states at each Co site. 
At Co$^{2+}$ sites, $3d$ orbitals are aligned along the $a$-axis 
while at Co$^{3+}$ sites, along the $b$-axis.
}
\end{figure}

In Fig. \ref{fig4}, we have plotted the geometry of the orbital ordering 
which is obtained from the orbital dependent occupancy of 
the $3d$ minority-spin states at each Co site. 
At Co$^{2+}$ sites, the orbitals are aligned along $a$-axis,
while at Co$^{3+}$ sites, the orbitals are along $b$-axis.
This feature is understandable by considering that the bond length 
of $d$(Co$^{2+}$-O) is larger in $a$-axis than in $b$-axis 
and {\it vice versa} for that of $d$(Co$^{3+}$-O). 
As for the charge ordering configuration, Co$^{2+}$ 
and Co$^{3+}$ chains of a stripe type are formed in the $ab$ plane 
along $b$-axis, 
which are alternating in the $a$ and $c$ direction\cite{Vogt,Suard}.
This is in contrast to the charge ordering observed in the isostructural
YBaMn$_2$O$_5$. In YBaMn$_2$O$_5$, the Mn$^{2+}$/Mn$^{3+}$ orders in a 
checkerboard  type\cite{Millange}. This difference gives rise to the
different magnetic structures: 
$G$-type AFM phase for YBaCo$_2$O$_5$ and $G$-type ferrimagnetic phase
for YBaMn$_2$O$_5$.
The theoretical result of the charge and orbital ordering geometry coincides 
with the experimentally proposed one\cite{Vogt,Suard}.
Thus, it can be inferred that the deformed bond lengths of $d$(Co-O) 
determine the charge and orbital ordering geometry.

The $G$-type AFM ordering in YBaCo$_2$O$_5$ is consistent 
with the above charge and orbital ordering geometry. 
In the Co$^{3+}$ chains, the kinetic-exchange energy gain 
between the occupied $d_{yz}$ and $d_{zx}$ states 
and the empty $d_{x^2-y^2}$ state for the AFM configuration 
of neighboring sites stabilizes the AFM spin ordering.
In a similar way, the AFM ordering between neighboring 
Co$^{3+}$ and Co$^{2+}$ ions can be explained.
The AFM ordering in the Co$^{2+}$ chains, however, is hard to understand 
in terms of the direct kinetic-exchange gain, 
because the overlap integral between two neighboring Co$^{2+}$ ions 
would be negligible as seen in Fig. \ref{fig4}. 
Instead, the AFM interaction in the Co$^{2+}$ chains is expected  
to be derived indirectly via the Co$^{2+}$-Co$^{3+}$ 
and Co$^{3+}$-Co$^{3+}$ AFM interactions.

As mentioned above, the structural transition plays a crucial role 
in determining  the ground state properties of YBaCo$_2$O$_5$.
Although the tetragonal to orthorhombic structural transition occurs 
simultaneously with the $G$-type AFM transition at $T_{\rm N} \sim 330$ K,
the structural deformation is not significant above 220 K.
Only near $T_{\rm CO} \sim 220$ K, the lattice splitting 
between $a$- and $b$-axis becomes pronounced and 
the charge and orbital ordering emerges
with the Co$^{2+}$ spin state transition from the low to high spin.
Furthermore, it is known that the long-range charge ordering and the
spin state transition are very sensitive to the oxygen stoichiometry\cite{Vogt}.
Therefore, the structural distortion is thought 
to be responsible for the orderings and the spin state transition 
by inducing different local environment at each Co ion site.
This feature implies that the electron-lattice interaction is very
important in this system.
A detailed study on the electron-phonon interaction effects 
in YBaCo$_2$O$_5$ is urgently demanded.   

In conclusion, we have performed the LSDA + $U$ calculations 
for a new spin-charge-orbital-lattice coupled system YBaCo$_2$O$_5$.
It is found that the Co$^{2+}$/Co$^{3+}$ charge and orbital ordering 
and the Co$^{2+}$ HS state transition are closely correlated with  
the lattice distortion from the tetragonal to orthorhombic structure. 
The orbital moment has a substantially large contribution to the 
total magnetic moment.
All of the effects of the Coulomb correlation, the spin-orbit coupling,
and the electron-phonon interaction should be properly taken into account 
to understand physical properties of YBaCo$_2$O$_5$. 

Acknowledgements$-$
This work was supported by the KOSEF (1999-2-114-002-5)
and by the Brain Korea 21 Project.


\begin{thebibliography}{30}
\bibitem{Nominal} The valency assignments are nominal, 
         because  the number of occupied $d$-electrons 
         inside the atomic sphere of Co is not an integer 
         due to the hybridization with other bands (see Table I).
\end{thebibliography}
\end{document}